2025

# Computationally Intensive Research: Advancing a Role for Secondary Analysis of Qualitative Data

Kaveh Mohajeri
*IESEG School of Management*, k.mohajeri@ieseg.fr

Amir Karami
*University of Alabama at Birmingham*, karami@uab.edu

# Computationally Intensive Research: Advancing a Role for Secondary Analysis of Qualitative Data

**Kaveh Mohajeri,[1] Amir Karami[2]**

[1]IESEG School of Management, France, k.mohajeri@ieseg.fr
[2] Collat School of Business, University of Alabama at Birmingham, USA, karami@uab.edu

## Abstract

This paper draws attention to the potential of computational methods in reworking data generated in past qualitative studies. While qualitative inquiries often produce rich data through rigorous and resource-intensive processes, much of this data usually remains unused. In this paper, we first make a general case for secondary analysis of qualitative data by discussing its benefits, distinctions, and epistemological aspects. We then argue for opportunities with computationally intensive secondary analysis, highlighting the possibility of drawing on data assemblages spanning multiple contexts and timeframes to address cross-contextual and longitudinal research phenomena and questions. We propose a scheme to perform computationally intensive secondary analysis and advance ideas on how this approach can help facilitate the development of innovative research designs. Finally, we enumerate some key challenges and ongoing concerns associated with qualitative data sharing and reuse.

## 1 Introduction

In their *MIS Quarterly* editorial on computationally intensive research, Miranda et al. (2022) aptly observe that "method choices are only loosely coupled to the [type of] data" (p. viii). Numerous studies within the IS field and beyond demonstrate that computational methods can effectively analyze various types and formats of data, including quantitative and qualitative trace data drawn from digital environments, census data, textual documents, images, and videos (e.g., Bahmanyar et al., 2018; Goldberg, 2011; Lindberg et al., 2016). Still, computationally intensive research dealing with qualitative data often focuses on "found" data, such as *digital trace data* (Berente et al., 2019; Howison et al., 2011; Sarkar, 2021), academic publications (e.g., Larsen et al., 2008; Mortenson & Vidgen, 2016), and organizational documents or reports (e.g., Harrison et al., 2019; Huang et al., 2018). While some studies have long hinted at the potential for computational analysis of data generated through qualitative studies (e.g., Indulska et al., 2012), few, if any, have pursued this approach, at least as far as the IS scholarship is concerned. Furthermore, the epistemological and methodological issues of applying computational methods in this context remain underexplored.

Data generated through qualitative studies (e.g., interview transcripts, open-ended survey responses, field notes, and research diaries) are precious materials, typically produced through rigorous, resource-intensive processes and usually accompanied by documented inquiry methods. However, qualitative researchers frequently note that much of their laboriously generated data can remain unused and that only a portion is often the subject of final analysis and publication (Davidson et al., 2019; Fielding & Fielding, 2000). Likewise, in the IS field, we are long aware that, for instance, the abundance of case studies represents a largely untapped pool of empirical evidence (Myers & Avison, 1997). Despite this, researchers may often not be at ease utilizing computational methods to help streamline their engagements with data generated through qualitative research. Due to specific epistemological justifications and traditions, committed qualitative researchers may still favor conventional, "manual" analysis when it comes to "deep data" research involving primary data (Davidson et al., 2019; Bruns, 2013). Concerns also exist about the applicability of computational methods. Most notably, logistical constraints in qualitative inquiries may cause primary data corpora to end up lacking the scale, if not other properties, required for computational methods to yield reliable results, especially with more complex algorithms (Izonin & Tkachenko, 2022; Schmiedel et al., 2019; van Loon, 2022).

This Research Perspectives paper aims to draw attention to—and contribute to a scholarly conversation about—the potential of leveraging computational methods in analyzing data generated through qualitative studies and how it can be a valuable addition to the practice of computationally intensive research. While computational methods can legitimately be worked out in certain conditions involving primary qualitative data, our focus in this paper is on opportunities with *secondary analysis of qualitative data* (hereafter, SAQD)—the computational reworking of data generated in previous qualitative studies, independent of and transcending the theoretical interests or even scope of those studies. We begin by making a general case for SAQD in IS research. Next, we argue for *computationally intensive SAQD* (hereafter, CI-SAQD), contending that even more benefits can be gained from SAQD when computational techniques augment it. This is particularly true when moving beyond an individual data set to analyze an assemblage of data sets spanning multiple contexts and timeframes to theorize for cross-contextual or longitudinal issues that are beyond the foci of each constituent data set individually (Davidson et al., 2019). We provide an overview of some relevant computational methods and propose a specific scheme to conduct CI-SAQD. In addition, we explore how CI-SAQD can facilitate the development of innovative research designs. Finally, we briefly discuss some key ongoing challenges and limitations of research that relies on SAQD.

## 2   Making a Case for SAQD in IS Research

For much of the twentieth century, data reuse in social science research was primarily associated with quantitative data sets (Fielding, 2000; Goodwin, 2012; Heaton, 2004). Ideas and practices around qualitative data reuse were rare, with notable exceptions, such as

Barney Glaser's articles in the 1960s (Glaser, 1962, 1963). By the late 1990s, however, the landscape began to shift considerably, particularly in North America and the United Kingdom, where researchers increasingly recognized the significance of the "secondary analysis" of qualitative data previously generated through rigorous research designs and structuring (Heaton, 1998, 2004).

### 2.1   SAQD: Benefits and Distinctions

Although data generated in qualitative studies represents a vastly underutilized resource, many IS researchers may still question or be unaware of the benefits of using such data for new, secondary studies (see Skinner et al., 2022). In general, the significance of qualitative data sharing and its corollary, SAQD, have been widely addressed based on several different grounds (e.g., Fielding & Fielding, 2000; Goodwin, 2012; Hammersley, 1997; Heaton, 2004; Hinds et al., 1997; Hughes et al., 2020; Mannheimer et al., 2019; Moore, 2007). In the following paragraphs, we discuss three primary arguments behind advocating SAQD. Also, we differentiate SAQD from other well-known approaches, which involve analyzing secondary qualitative data or reviewing previous qualitative research.

The first primary argument for SAQD highlights cases where collecting primary data is overly cumbersome, if not infeasible or unjustifiable, yet reliance on data from qualitative fieldwork remains indispensable. Research on sensitive topics, hard-to-reach populations, or past events often makes primary data collection unreasonably difficult. There are also situations where researchers aim to avoid overburdening the informants (Fielding, 2000; Heaton, 2004; Long-Sutehall et al., 2011; Mannheimer et al., 2019; Sandelowski, 1997). For instance, Bishop and Kuula-Luumi (2017) point to the health area as an incredibly fertile domain in terms of the range of sensitive topics with general appeal that have been approached through various genres of SAQD by using data from repositories such as the one provided and curated by healthtalk.org. One specific example in IS research that resonates well with this type of argument for SAQD relates to the research stream on algorithmically managed work. The literature shows that recruiting interviewees in specific algorithmically managed settings can face notable challenges. For instance, Tarafdar et al. (2022) explain how it is complicated to interview drivers working for ridesharing platforms (e.g., Uber, Lyft) as they often lack time due to demanding schedules and are wary of being interviewed because of the frequent negative press coverage of ridesharing companies. In such cases, we believe data collected in earlier studies would constitute a highly precious source of evidence worth systematically retaining and sharing for future secondary analysis.

The second type of argument to pursue SAQD centers on the economy and sustainability of qualitative research. Collecting primary data can be "very resource-intensive and beyond the means of most social scientists who do not have access to significant sources of research funding" (Goodwin, 2012, p. xxii). There are also broader recognitions, maintaining that data sharing and reuse is particularly beneficial to encourage new research partnerships, promote research transparency, provide resources for student research, and "maximize the payoff of public investments in research and education" while requiring less burden on research subjects (Mannheimer et al., 2019, p. 644; Heaton, 2004; Bishop & Kuula-Luumi, 2017).

The third type of argument goes beyond logistical or practical concerns to underscore the substantive benefits of SAQD. Advocates argue that reusing qualitative data can foster new research questions/designs and enable innovative research with novel and impactful findings. As early as the 1960s, Barney Glaser (1963) noted that secondary analysis by independent researchers "can lend new strength to the body of fundamental social knowledge" (p. 11). Subsequent scholars have also similarly emphasized SAQD's potential (Fielding & Fielding, 2000; Hammersley, 1997; Heaton, 2004). Heaton (2004), for instance, suggests that SAQD "epitomizes the flexible character of qualitative research, enabling researchers to find innovative ways of using pre-existing data" (p. 71). She also explains that SAQD can be beneficial in *salvaging* data from primary work to address new/additional research questions or for research verification or refinement purposes. Fielding and Fielding (2000), in the sociology domain, showcase the potential of such a practice. They revisited a seminal study on prison life conducted by Cohen and Taylor (1972), demonstrating "support for an alternative, if complementary, conceptualisation, using archived data from the original study" (Fielding & Fielding, 2000, p. 671). Likewise, Weick et al. (2005) practice SAQD in the organizational studies domain, leveraging interview data from a clinical nursing study to illustrate key elements of their influential sensemaking framework. Data from multiple previous studies dealing with the same (or similar) population(s) may also be systematically leveraged to evaluate the generalizability of primary research findings. This can thus facilitate addressing what may be frequently perceived as a weakness of qualitative research (Hammersley, 1997). In addition, Hammersley (1997) draws attention to how, in general, qualitative data sharing and SAQD can open possibilities of wide-ranging comparative analysis.

These substantive benefits of SAQD continue to be echoed in more recent scholarship across disciplines such as political science (e.g., Elman et al., 2010; Kern & Mustasilta, 2023), health (e.g., Chatfield, 2020; Tate & Happ, 2018), and the social sciences (e.g., Courage, 2019). However, most notably, SAQD is increasingly valued for its potential to support longitudinal and cross-contextual theorizing. Scholars emphasize the possibility of "scaling up"[1] across qualitative data sets from multiple studies that extend over different time frames and contexts (Davidson et al., 2019; Edwards et al., 2021; Mason, 2002). For instance, the Timescapes initiative (2007-2012), funded by the UK's Economic and Social Research Council, demonstrated how longitudinal qualitative data archives can enable groundbreaking temporal research to explore "the lived experience of change and continuity in the social world" (Neale et al., 2012, p. 5).

In IS research, leveraging the richness of multiple qualitative studies to drive innovative theorizing and uncover novel perspectives is already an established practice, though through approaches other than SAQD (e.g., Rivard & Lapointe, 2012; Stafford & Farshadkhah, 2020; Tana et al., 2023). To clarify SAQD's distinctiveness, it is thus helpful to briefly compare it to other prominent approaches, specifically qualitative meta-synthesis (Hoon, 2013; Zimmer, 2006) and the case survey method (Larsson, 1993; Lucas, 1974).

SAQD is distinct in that it mainly relies directly on preexisting data rather than on research findings. This largely sets SAQD apart from approaches in the "review research" tradition, such as meta-analysis and meta-synthesis (Kunisch et al., 2023), where the researcher often "analyzes the analyses" or "codes the codes" (Stafford & Farshadkhah, 2020) and seldom engages directly with the data itself (Heaton, 1998, 2004; Zimmer, 2006). Furthermore, qualitative meta-synthesis (and most other types of "review research") typically requires substantial topic homogeneity between the studies being synthesized and the meta-synthesis itself (Zimmer, 2006). By contrast, SAQD can completely transcend and be independent of the theoretical interests or scopes of the primary studies, as exemplified in the *supra-analysis* type of SAQD (Heaton, 2004). Finally, SAQD studies primarily entail *qualitative* data analysis, distinguishing them from case survey studies. With the understandings and explanations' (Mason, 2002, p. 4)." However, Davidson et al. (2019, p. 366) emphasize the need to move beyond "scaling up," as the term does not fully reflect the iterative process of working with assemblages of data sets, where "not only are data put together but they may be organised in new ways."

---

[1] Davidson et al. (2019, p. 364) note: "The case for what was referred to as 'scaling up' across multiple qualitative data sets saw light in a scoping paper produced for the UK's Economic and Social Research Council by Jennifer Mason in 2002, which concluded: 'perhaps the most significant opportunity offered to qualitative data is the possibility of 'scaling up' through data sharing, to produce cross-contextual

case survey method, researchers typically aim to study "many issues in many cases" (Larsson, 1993, p. 1515) through an essentially *quantitative* review of qualitative data (Jurisch et al., 2013), where the researcher follows a coding procedure of assigning numbers to different properties/features of the constituent case studies.

## 2.2 Epistemological Considerations

SAQD has sparked intense debates regarding the epistemology of qualitative research for over two decades (Vila-Henninger et al., 2022; Hammersley, 1997, 2010; Savage, 2005). The unease and debates around SAQD often reflect broader tensions between different qualitative research traditions or, more generally, between qualitative and quantitative approaches to social science research (Heaton, 2004; Bishop, 2007). For instance, the primary-versus-secondary debate surrounding SAQD has been described as "a proxy for other debates: positivism/interactionism, realism/post-modernism, subjectivity/authorial authority, and even academic freedom/neo-managerialism" (Bishop, 2007, p. 53; also see Moore, 2005).

Following Heaton (2004), we believe these epistemological debates may be best addressed by first examining different views on the relationship between qualitative and quantitative research. Bryman (1988) and Hammersley (1996) provide useful frameworks in this regard. Bryman (1988) distinguishes between "epistemological" and "technical" perspectives, which align closely with Hammersley's (1996) "paradigm loyalty" versus "methodological eclecticism" positions. The "epistemological" or "paradigm loyalty" view sees quantitative and qualitative research as fundamentally incompatible due to their *incommensurable* ontological and epistemological beliefs. In contrast, the "technical" or "methodological eclecticism" view emphasizes the practical rather than philosophical aspects of social inquiry. It considers the two quantitative and qualitative approaches complementary, having different strengths and weaknesses, and thus suitable for investigating different kinds of research questions. We observe that the latter view has become increasingly prevalent among IS scholars, who often embrace qualitative research as a broad church encompassing various paradigms (e.g., positivist, interpretive, critical, postmodern, etc.) and genres (Cecez-Kecmanovic & Kennan, 2013; Sarker et al., 2018).

The distinction between the two views outlined above would be instrumental in understanding the two main sides of the debates over SAQD. Proponents often align with the "technical" or "methodological eclecticism" view, which Sarker et al. (2018) associate with a data-centric approach. Skeptics, however, tend to lean toward the "epistemological" or "paradigm loyalty" view, aligning more with an interpretation-centric approach (Sarker et al., 2018). Still, a closer examination needs to address the two sides' positions regarding two fundamental issues: data "fit" and the issue of "not being there" (Heaton, 2004; Hughes et al., 2020; Vila-Henninger et al., 2022). Data "fit" concerns whether data from previous qualitative studies can be legitimately reused for new research purposes. The "not being there" issue raises questions about how the secondary researcher's distance from the inquiry context might affect data interpretation.

On the issue of data "fit," critics of SAQD often question the practice of treating qualitative data as a "reified neutral product" that can legitimately be divorced from the situated dynamics of its generation context and repurposed for secondary research (Hughes et al., 2020, p. 567; also, see Moore, 2007). Proponents, however, take a pragmatic stance, in line with the view of data as "representative facts or shared reality" (Sarker et al., 2018), thus supporting qualitative data reuse. The SAQD literature also suggests that data "fit" is not fixed but depends on factors such as the extent of missing data in the corpus, the alignment between primary and secondary research questions, and the suitability of the data's nature, structure, and format for the intended secondary analysis methods (Hinds et al., 1997; Thorne, 1994).

Regarding the second issue, critics argue that meaningful qualitative analysis relies heavily on researchers' prolonged, immersive contact with the field, a result of "being there" during data collection (Hughes et al., 2020, p. 566; also, see Mauthner & Parry, 2009, 2013). Proponents, however, tend to "challenge the notion that 'remove' from the original spatial, temporal and epistemic context of the production of 'primary' data is *exclusively* a form of deficit" (Hughes et al., 2020, p. 567, emphasis original; also see Irwin & Winterton, 2011). In other words, the logic here is that opportunities for insight are not inextricably tied to the presence in the immediate data generation contexts; instead, certain observations become possible "precisely when we are not there 'at the moment'" (Hughes et al., 2020, p. 567).

## 2.3 The Significance of SAQD for the IS Scholarship

Against the backdrop laid out above, we contend that SAQD should first and foremost be embraced in the IS field for its great potential to problematize long-held assumptions about qualitative inquiry, data, and the relationship between researchers and data. SAQD allows IS qualitative researchers to move beyond the traditional view that only what researchers produce as data through direct sensory engagement in specific social contexts is valuable. Instead, it highlights the further significance of what such data "when treated as particular kinds of evidence through specific forms of research engagement and apprehension can be used to say about the social world" (Hughes et al., 2020, p. 567).

By reworking preexisting qualitative data, IS scholars can expand the contributions of past studies from merely their findings to what can be additionally unlocked through the reuse of their rich data sets (cf. Glaser, 1962). SAQD would also motivate IS researchers to adopt a *bricoleur* mindset (Heaton, 2004; cf. Denzin & Lincoln, 2011), leveraging diverse theoretical and analytical approaches while mining a wide array of qualitative data sets to explore novel or supplementary research questions. This completely aligns with the recurrent calls for more flexible, eclectic, and innovative research styles within IS and beyond.

Furthermore, it is encouraging that the epistemological and methodological concerns around qualitative data reuse have long begun to be addressed through various research practices and strategies. For instance, to ensure data "fit," secondary studies can draw on researchers' prior familiarity with the data set. Another practice can be moving away from treating qualitative data as "given" and adapting it to match the goals of the secondary analysis. Finally, a third practice would be augmenting the data set with additional primary data (see Heaton, 2004). In addition, to address the issue of "not being there," one strategy is to deposit all relevant study documentation (e.g., field notes, research diaries, and other correspondence related to research execution) by primary researchers when archiving qualitative data (Corti & Thompson, 1998; Fink, 2000; Mannheimer et al., 2019). Another strategy is to consult with primary researchers, where possible, to gain deeper insights into the original work and become sensitized to the context of the primary study (Hinds et al., 1997).

# 3  Toward Computationally Intensive SAQD

Computational analysis of qualitative data has substantially impacted management and IS scholarships, enabling fresh approaches to theorizing and reexamining previously intricate problems and questions, especially with large-scale [2] data sets (Hannigan et al., 2019; Miranda et al., 2022). However, as stated earlier, computational methods are predominantly utilized for analyzing "found" qualitative data. This tendency can even be observed in the methodological thinking and texts that aim to justify computational methods (e.g., Schmiedel et al., 2019; Berente et al., 2019), where these methods are often positioned in contrast to or separate from "conventional" or "traditional" qualitative research that relies on, for instance, interview data. We posit that this division is unnecessarily restrictive, if not damaging. Data generated through qualitative studies, as with digital trace data and other types of "found" data, can yield valuable insights and help ambitious and innovative theorizing when analyzed computationally. This is where we introduce CI-SAQD: the integration of computational methods with secondary analysis of qualitative data.

While computational methods could be applied to primary qualitative data, integrating them with SAQD has unique advantages. SAQD enables the assembly of data sets from multiple studies, addressing the need for larger data corpora, which is often required for reliable computational analysis (Izonin & Tkachenko, 2022; van Loon, 2022). In addition, CI-SAQD can allow longitudinal and cross-contextual theorizing even more effectively than conventional SAQD by computationally engaging with assemblages of multiple data sets spanning different time frames and contexts. This, in turn, can significantly enhance research richness and help gain substantive insights into social processes (Davidson et al., 2019). CI-SAQD can also substantially facilitate certain research designs, incorporating multiple and various modes of theorizing with different data sets within a single study.

The following subsections provide an overview of computational methods relevant to CI-SAQD, propose a specific CI-SAQD scheme, and expound on how CI-SAQD can enable innovative research designs.

## 3.1  Overview of Relevant Computational Methods

Computational methods generally handle two types of data: structured (organized and stored in tabular formats with rows and columns representing instances and features) and unstructured (often lacking a standard shape or organization). Most qualitative data, such as interview transcripts, are unstructured, for which unsupervised[3] computational techniques are commonly used to derive structured understandings and descriptions. In particular, *topic modeling* has been widely applied across various domains (e.g., Bybee et al., 2023; Greve et al., 2022; Karami et al., 2020b; Grisham et al., 2023). Topic models often rely on a particular computational foundation, such as linear algebra, probability, neural networks, or fuzzy clustering (Karami et al., 2020a; Abdelrazek et al.,

---

[2] Schmiedel et al. (2019, p. 946) provide a helpful perspective on the issue of data corpus size in computational analysis: "Unfortunately, existing literature to date lacks theoretically justified guidelines regarding minimal corpus size, but insights from empirical studies can provide some guidelines. Experimental studies suggest that the results of LDA for corpora with few documents (i.e., <100) are very difficult to

interpret, even if the documents are long; the interpretability of topic models improves with increased corpus size and stabilizes at around 1,000 documents (Nguyen, 2015)."

[3] Unsupervised techniques organize and work with "unlabeled" data to detect patterns for certain purposes, such as clustering (e.g., clustering customers based on their online shopping behavior) (Shah, 2020).

2023). Four specific topic models are briefly reviewed below.

Latent Semantic Analysis (LSA) is a topic model that draws on linear algebra using Singular Value Decomposition (SVD) to reduce the dimensionality of term-document matrices, uncovering hidden semantic structures in textual data (Deerwester et al., 1990). It has been applied in several domains, such as political science (e.g., Valdez et al., 2018), business (e.g., Lau et al., 2014), and health (e.g., Han & Choi, 2010). However, LSA is believed to struggle with estimating the number of topics (or dimensions) and assigning topics to new, unseen documents, limiting its applicability (Blei et al., 2003; Abdelrazek et al., 2023; Zengul et al., 2023).

Latent Dirichlet Allocation (LDA) is a probabilistic topic model, where each piece of unstructured data, such as a transcript, is assumed to contain multiple topics, each representing a collection of semantically related words. For instance, LDA might categorize terms like "data," "number," and "computer" under a topic, which can be labeled as "data analysis" (Blei, 2012; Blei et al., 2003; Boyd-Graber et al., 2017). LDA is widely applied to data sets of various scales across domains like management, health, and politics (e.g., Boyd-Graber et al., 2017; Hannigan et al., 2019). It substantially facilitates assigning topics to new documents, enabling more straightforward interpretation and offering better performance than earlier models like LSA. However, LDA requires careful parameter tuning and can involve substantial human effort for interpretation (Rijcken et al., 2021).

Fuzzy Latent Semantic Analysis (FLSA), based on fuzzy clustering, allows each data item to be assigned to multiple clusters (Karami et al., 2018). FLSA functions based on degrees of truth rather than traditional binary values, where each keyword is associated with each document to a certain fuzzy membership degree. FLSA has primarily been used in health research (Abdelrazek et al., 2023). One main advantage of this topic model is to address redundancy issues, typically offering higher topic coherence than LDA (Rijcken et al., 2021). However, FLSA has yet to be widely tested with large-scale data sets.

Top2Vec uses the Word2Vec neural network to detect word similarity within documents (Angelov, 2020). It is effective for large-scale data sets and has been applied in areas like analyzing customer reviews (Ozansoy Çadırcı, 2023) and identifying patient needs (Karas et al., 2022). Despite its strengths, Top2Vec has limitations, including challenges with parameter interpretation (Abdelrazek et al., 2023), excessive topic generation, and suboptimal performance with smaller data sets (Zengul et al., 2023). Additionally, while documents can theoretically be linked with multiple topics, Top2Vec typically assigns only one dominant topic per document (Egger & Yu, 2022).

To sum up, when selecting a topic modeling technique, researchers should consider three critical factors: data set scale (Egger & Yu, 2022; Zengul et al., 2023; Abdelrazek et al., 2023), ease of parameter and topic interpretation (Zengul et al., 2023; Abdelrazek et al., 2023), and topic quality and coherence (Karami, 2015; Rijcken et al., 2021).

## 3.2 Computationally Intensive SAQD: A Proposed Scheme

In this subsection, we propose a scheme to conduct CI-SAQD, using LDA as the underlying computational technique. Several reasons underpin this choice. First, LDA allows topic numbers to be determined through both quantitative (e.g., coherence analysis) and qualitative (e.g., human coding) methods. While models like Top2Vec can automate this process, they risk overfitting by generating an excessive number of topics (Zengul et al., 2023). Second, probabilistic topic models like LDA were designed to address certain limitations of linear algebra-based methods, such as the Latent Categorization Method (Larsen et al., 2008), particularly in terms of the inference of document-topic distributions (Blei et al., 2003). Compared to such methods, LDA consistently produces more coherent topics (Egger & Yu, 2022; Zengul et al., 2023). Another advantage of LDA is its widespread accessibility and popularity (Egger & Yu, 2022). It is available in various programming languages (e.g., C, Java, Python, and R), making it usable by researchers with diverse technical backgrounds. LDA's methodological alignment with the grounded theory approach further enhances its appeal (Baumer et al., 2017). Furthermore, LDA's performance has been tested across varying data set scales, demonstrating superior results with larger data corpora (Schmiedel et al., 2019; Tang et al., 2014). When applied to large data sets, LDA can afford applications such as tracking topic evolution over time and space, assessing topic associations with external factors like financial indices, conducting topic comparisons across multiple data sets, and uncovering novel semantic patterns that may redefine research questions (Hannigan et al., 2019; Kiley et al., 2023; Schmiedel et al., 2019). Finally, LDA is well-established in the IS field, where it has been recognized for its potential to be integrated with conventional qualitative methods for theory building and to address a variety of research problems across different contexts (Gjerstad et al., 2021; Jung & Suh, 2019; Lappas et al., 2016; Rai, 2016; Yang & Subramanyam, 2023).

Figure 1 illustrates our proposed scheme, *one possible approach* (out of many) to integrating SAQD with computational elements and steps. The scheme focuses solely on the empirical phase of CI-SAQD research, which, like other types of research, is guided in the first place by the researcher's phenomena of interest, theoretical considerations, and research questions.

The empirical phase begins with constructing a corpus of qualitative data. Data sources may include published qualitative studies, archives containing data from previous qualitative studies, or even primary data from the researcher's past projects, according to Heaton (2004, 2008). The "data search & selection" step requires considering what is practically accessible as data, which is also contingent on the researcher's skills and resources. Research questions, epistemological stance, theorizing modes (e.g., inductive, deductive, abductive), units of analysis, and specific contextual preferences such as geography, time, or language would also fundamentally drive the broad and in-depth search processes for data.

During the broad search for data, researchers often make preliminary decisions about inclusion or exclusion in relation to the data corpus they assemble. Metadata, typically attached to archived data items, can help structure files in the assembled corpus (Davidson et al., 2019). The subsequent in-depth evaluation of "fit" between the research study criteria and provisionally selected data sets is a more challenging task. Based on the SAQD literature (e.g., Fielding, 2000; Hammersley, 1997, 2010; Heaton, 2004; Hinds et al., 1997), the notion of fit can be characterized by two dimensions: *suitability* and *sufficiency*.

Suitability refers to the data set's alignment with the researcher's phenomena of interest and ability to support the researcher's desired theorizing mode (e.g., inductive or deductive). Researchers may also draw from classic strategies for sampling and case selection (see Flyvbjerg, 2006). However, suitability often hinges on practical circumstances that may diverge from initial research expectations. Sufficiency, on the other hand, pertains to the richness and quality of the data set. Preexisting data must offer enough depth and detail to support the researcher's envisioned theorizing tasks. Richness can usually be gauged by examining the presence of "thick" descriptions or narratives that provide a chain of evidence, explaining sequences of unfolding events (Beaudry & Pinsonneault, 2005; Paré & Elam, 1997; Yin, 2014). Data quality can also be assessed by investigating the primary studies' rigor. According to Hinds et al. (1997), such an investigation may involve examining the primary researchers' credentials, methodological expertise, and other factors, such as the amount of time spent at the site for data collection.

The "pre-processing" step aims to clean data and extract features from the data corpus constructed in the previous step. Pre-processing involves removing irrelevant content (e.g., stop words), normalizing related terms (e.g., lowercase conversion), and enhancing semantic information capture (e.g., handling negation) (Hickman et al., 2022). Feature extraction methods, such as the Bag-of-Words (BoW) model (Aggarwal, 2015), are also applied at this stage.

In the "semantic exploration" step, text data is analyzed using LDA. This requires setting three specific parameters. The first one is the number of topics. There are some methods to estimate the number of topics, such as *coherence analysis*, which measures semantic similarity among top words in a topic (Röder et al., 2015). The other two parameters are alpha and beta, representing document-topic and topic-word densities. Assuming each document represents a few topics, a small value often must be set for alpha. Also, beta is adjusted to a lower value when the identified topics have sparse word usage. Parameter tuning can thus significantly help avoid very general and shallow topics (Jian et al., 2014). Regarding semantic identification and analysis, one must consider what LDA produces as two key outputs: (1) the probability of each word per topic or P(W|T), essential for identifying topics, and (2) the probability of each topic per document or P(T|D), used to assess topic significance and enable further analyses (e.g., t-test or ANOVA on topic weights). Screening the identified topics may reveal opportunities for further refinements, such as adding more stop words to filter out common but semantically insignificant words. This suggests one must often iterate between "semantic exploration" and "pre-processing" to improve topic quality.

The "interpretation & visualization" step involves interpreting and labeling the topics identified and analyzed in the previous step. Researchers would perform tasks such as reviewing top words within topics (e.g., top 10 ranked by P(W|T)) and identifying relevant documents by sorting P(T|D). Researchers in a team may also employ methods such as *consensus coding* (Lim et al., 2015) to achieve agreement on topic meanings and labeling and group topics into higher-order categories. In addition, external auditors can validate these interpretation and labeling processes, adding an extra layer of rigor. Visualizations, such as word clouds, bar charts, line charts, and maps, can be used to illustrate findings, highlight topic dynamics, and explore patterns over time or space (Bai et al., 2021; Bennett et al., 2021; Lin et al., 2020). Finally, iterations between the "semantic exploration" and "interpretation & visualization" steps are often necessary to address problems such as unclear or overly general topics. Relying on relevant domain knowledge, researchers can fine-tune the number of topics, alpha and beta, to control topic diversity and word selection within topics (Jian et al., 2014; Wallach et al., 2009).

Epistemological discussions about what has been framed in various ways, including "the digitalization of qualitative research" (Lee & Sarker, 2023), also intersect with this "interpretation & visualization" process. Critics argue that "big data" and computational techniques can separate methods from methodology and discipline (Smith, 2014). However, scholars like Davidson et al. (2019) emphasize how computational approaches make "the shift between breadth and depth more transparent, enabling us to move across disciplinary and epistemological perspectives and introduce cross-contextual generalisations" (p. 373). Günther et al. (2023) describe this as a "reflexive dance" between researchers and algorithms, where the interplay—if researchers maintain an active and reflexive stance—can ensure enhancing transparency and meaning rather than introducing bias and opacity. We thus concur with the perspective that computational methods like topic modeling do not "spit out" answers. They support, rather than replace, researchers whose interpretive capabilities and decisions remain central to connecting results with research studies' broader theoretical framing and questions (Kiley et al., 2023).

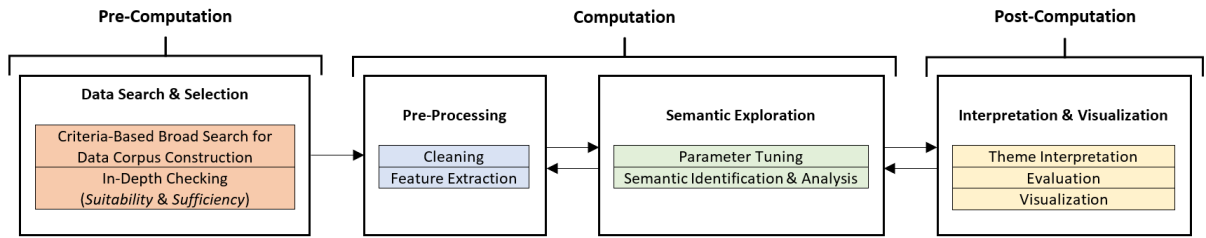

**Figure 1. A Scheme for Conducting Computationally Intensive SAQD**

**Table 1. A Summary of the Proposed CI-SAQD Scheme**

| Steps | Key issues and considerations | Examples of corresponding literature |
|---|---|---|
| Data search & selection | • Accessibility of preexisting data<br>• Use of qualitative data archives<br>• "Fit" between the research study criteria and provisionally selected data sets<br>• Suitability and sufficiency of preexisting data | Bishop & Kuula-Luumi (2017)<br>Davidson et al. (2019)<br>Hammersley (1997, 2010)<br>Heaton (2004)<br>Hinds et al. (1997) |
| Pre-processing | • Removing irrelevant content<br>• Normalizing related terms<br>• Enhancing semantic information capture<br>• Feature extraction methods | Aggarwal (2015)<br>Hickman et al. (2022) |
| Semantic exploration | • Setting the number of topics<br>• Setting alpha and beta<br>• P(W|T) and P(T|D) for semantic identification and analysis<br>• Iteration between the semantic exploration and pre-processing steps | Jian et al. (2014)<br>Röder et al. (2015)<br>Schmiedel et al. (2019) |
| Interpretation & visualization | • Topic interpretation and labeling<br>• Use of *consensus coding*<br>• Involving external auditors<br>• Use of visualizations<br>• Iteration between the interpretation & visualization and semantic exploration steps | Bai et al. (2021)<br>Bennett et al. (2021)<br>Jian et al. (2014)<br>Lim et al. (2015)<br>Wallach et al. (2009) |

## 3.3 Computationally Intensive SAQD: Innovative Research Designs

Theorizing can be seen as a systematic process involving a dialectic between data and generalizations aimed at accounting for empirical observations (Timmermans & Tavory, 2012). Qualitative research within IS and beyond is famous for approaching such dialectic often inductively, producing "either a substantive or a formal theory through a heuristic process of abstraction" (Timmermans & Tavory, 2012, p. 169). While inductive theory-building approaches still dominate, deductive qualitative research focused on theory testing has also gained traction over the past several decades. This mainly stems from the recognition that relying solely on quantitative methods for theory testing can be inadequate, if not damaging, particularly when assessing theories involving causal claims, emergent longitudinal relationships, dynamic processes, or human intentions *in situ* (Løkke & Sørensen, 2014; Miller & Tsang, 2011).

While theory testing is especially crucial in fields like management and IS, theories often remain unchallenged post-development (Suddaby et al., 2011; Fisher & Aguinis, 2017), and concerns persist about "the overabundance of weak and *untested* theory" (Cronin et al., 2021, p. 667, emphasis added). A contributing factor in qualitative research is the historical convention that theory building and testing require separate data sets (Hyde, 2000), which has created "practical" barriers to pursuing deductive theorizing within IS and beyond (Bitektine, 2008; Dubé & Paré, 2003; Løkke & Sørensen, 2014).

The abovementioned situation inspires the potential for new research designs integrating CI-SAQD, allowing multiple rounds of theorizing, *each based on a distinct data corpus*. In this context, we expect that computational methods could enhance efficiency, if nothing else, when engaging with these potentially large-scale data sets across theorizing rounds. In addition, we know that computational techniques like topic modeling are readily well-suited for various theorizing modes, enabling researchers to navigate inductive, deductive, and abductive approaches effectively (Fligstein et al., 2017; Haans, 2019; Hannigan et al., 2019; Kaplan & Vakili, 2015).

Research designs, such as those illustrated in Table 2, can broaden IS researchers' choices to account for complex phenomena by enabling novel forms of longitudinal and/or cross-contextual research. *Integrated Inductive-Deductive Theorizing* (II-DT) enables researchers to incorporate both inductive and deductive phases into their projects. For instance, in the first phase, researchers might inductively develop a theory of digital transformation through a primary case study in the manufacturing sector.

The second phase would then involve deductive testing of the theory, involving computational analysis of an assemblage of data from previous qualitative studies in two other sectors, i.e., computationally intensive *amplified analysis*.[4] Another approach, *Sequential Deductive Theorizing*, allows multiple rounds of theory testing within a single research project. After each round, the theory is revised before commencing the next round. For example, researchers might test and refine a well-established theory about algorithmic control by drawing on data from three major technology companies. Each round would then involve computational analysis of data about a distinct technology company, incorporating prior qualitative studies, digital trace data, and corporate reports, i.e., computationally intensive *assorted analysis*.[5]

We can also envision more advanced designs that extend beyond the two basic approaches outlined above. For instance, to enhance rigor in theorizing, one could combine the II-DT and sequential designs, where a theory developed in the initial phase of II-DT undergoes sequential testing in later phases. This approach can be referred to as *Sequential II-DT*. Alternatively, a *Parallel II-DT* approach could be employed, involving parallel, independent theory assessments during the second phase of II-DT using two or more distinct corpora of preexisting qualitative data.

To conclude this subsection, we acknowledge the growing movement in qualitative research advocating that "abduction, rather than induction, should be the guiding principle of empirically based theory construction" (Timmermans & Tavory, 2012, p. 167). Abduction is also associated with arguments challenging the perceived dichotomy between induction and deduction (Goldberg, 2015). Notably, recent perspectives suggest that an abductive orientation can better leverage SAQD's defining features in theorizing, especially considering that much of the criticism against SAQD is anchored in an inductive epistemology (Deterding & Waters, 2018; Vila-Henninger et al., 2022). We expect this abductive orientation also to benefit the research designs discussed earlier. For instance, we envisage that a research design like II-DT can be plausibly reworked into an integrative abductive process. This aligns with Charles S. Peirce's assertion that "abduction is an integral process of the scientific method" (Timmermans & Tavory, 2012, p. 171), encompassing both induction and deduction (Vila-Henninger et al., 2022). We anticipate this shift will open new avenues for exploration and application with SAQD and CI-SAQD. Consequently, we call for more nuanced methodological treatments of the subject in the future.

---

[4] Amplified analysis is a type of SAQD where researchers combine or compare two or more data sets from previous studies. Of course, the data sets may belong to various research contexts, periods, and, or analysis levels (Heaton, 2004, 2008).

[5] Assorted analysis is a type of SAQD where researchers reuse qualitative data from previous studies in combination with primary or other types of qualitative data (Heaton, 2004, 2008).

**Table 2. Examples of Research Designs Augmented by CI-SAQD**

| Type of research design | Illustration of an example scenario | Description of an example scenario |
|---|---|---|
| Integrated inductive-deductive theorizing (II-DT) | 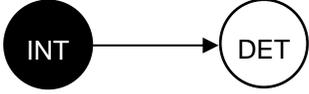 | The INT phase is accomplished through a conventional qualitative study using primary qualitative data. The DET phase entails *amplified analysis* to test the theory built in the INT phase. The researcher uses computational methods in the DET phase to engage with a corpus of preexisting qualitative data. |
| Sequential deductive theorizing | 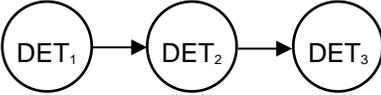 | Three consecutive rounds of DET with an already established theory are conducted. Each round of DET entails *assorted analysis*, where the researcher uses computational methods to engage with a distinct corpus of preexisting qualitative data. |
| Sequential II-DT | 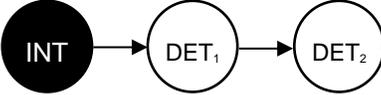 | The INT phase is accomplished through a conventional qualitative study that draws on primary qualitative data. Two consecutive DET rounds follow, each involving *amplified analysis* with computational methods applied to a distinct corpus of preexisting qualitative data. |
| Parallel II-DT | 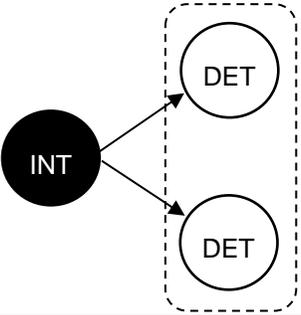 | The INT phase is accomplished through a qualitative study drawing on digital trace data. Two parallel DET rounds are conducted in the second phase, each involving *amplified analysis* with computational methods applied to a distinct corpus of preexisting qualitative data. |

*Note:* INT: inductive theorizing, DET: deductive theorizing)

# 4 SAQD: Challenges and Limitations

Practicing SAQD often involves two broad categories of challenges: practical challenges and deeper epistemological, ethical, and legal challenges. The practical challenges primarily revolve around the availability, suitability, and adequacy of preexisting qualitative data. SAQD becomes a viable strategy only when researchers can locate and access relevant and sufficient datasets. For instance, identifying suitable data from past qualitative studies may prove difficult if the phenomenon of interest is highly novel or unusual. In another case, researchers might find a potentially relevant dataset in published studies or archives;

however, further investigation might show that the dataset is unsuitable for the intended secondary analysis. Still, many of these practical challenges can be mitigated or resolved by adhering to specific guidelines, including those concerning the *suitability* and *sufficiency* of preexisting data we previously discussed.

The second category of challenges involves more complex issues and questions that remain the subject of ongoing debate across disciplines (e.g., Feldman & Shaw, 2019; Heaton, 2004; Mannheimer et al., 2019; Mauthner & Parry, 2009, 2013). For instance, ethical concerns about confidentiality agreements and ensuring participant anonymity are central to the discussions about preserving and sharing qualitative data (Neale, 2013; Ruggiano & Perry, 2019). Similarly, as stated, epistemological debates continue over whether qualitative data can plausibly be "re-used." At the same time, some argue that challenges like ensuring a proper fit between data and theory or avoiding data misinterpretation are not unique to SAQD and are hurdles faced in both primary and secondary qualitative studies (Fielding, 2000; Hughes et al., 2020; Moore, 2007). In addition, even some critics of SAQD concede that their reservations "need not entirely preclude support for data sharing and re-use or all opportunities for comparative analysis" (Feldman & Shaw, 2019, p. 15).

We believe that many challenges associated with SAQD can be addressed by developing robust protocols and structures to foster collaboration among such entities as researchers, data repositories, university libraries, and funding institutions (Feldman & Shaw, 2019; Mannheimer et al., 2019). Additionally, we emphasize the unique opportunity—and responsibility—for our community, especially in the era of computational methods, to engage more deeply with SAQD, advancing its various epistemological, methodological, and practical aspects. A first promising step forward can involve initiatives to establish and maintain digital archives. Such initiatives would promote a culture of preserving and sharing precious, rich qualitative data on contemporary and historical IS phenomena, serving current research needs and providing a crucial resource for future generations of IS scholars.

## 5 Conclusion

This paper aims to contribute to a scholarly conversation on leveraging computational methods when reusing data generated in previous qualitative studies. A similar theme has also recently gained momentum and echoed in the management and social sciences research communities (Davidson et al., 2019; Hannigan et al., 2019). Our work also aligns with broader recognitions that the digitalization of research resources and processes continues to have consequential impacts on qualitative research practices (e.g., see Simeonova &

Galliers, 2023). New mindsets and novel approaches have emerged, especially over the past two decades, regarding qualitative data generation, preservation, sharing, and analysis. Digital advancements have spurred the prevalence of qualitative data sharing and reuse, while the "archiving of qualitative research data is increasingly becoming a matter of national policy and practice in the United Kingdom, United States, Canada and Europe" (Mauthner & Parry, 2009, p. 301). In addition, disruptive innovations, particularly with generative artificial intelligence (AI), signal notable shifts in qualitative research concerning established methods and practices such as *coding*. However, admittedly, it still seems inconceivable that AI tools and language models serve as anything more than aids for generating inputs to the more interpretive aspects of qualitative research, remaining firmly within the realm of human expertise (Morgan, 2023; Perkins & Roe, 2024; Wachinger et al., 2024).

As researchers increasingly gain access to preexisting qualitative data—such as interview transcripts, field notes, and observational records—through digital archives and other mediums, the sheer scale of such data often exceeds the reading and analysis capacity of qualitative research teams. This paper's ideas and guidelines about computationally intensive secondary analysis not only offer a pathway to harness this data wealth more efficiently but also lay a groundwork for rigorous research drawing on carefully constructed assemblages of such data to innovatively approach today's intricate phenomena and questions.

## Acknowledgments

We sincerely thank the senior editor, Dirk Hovorka, and the reviewers for their invaluable guidance, constructive feedback, and patience. We also gratefully acknowledge the opportunity to present an earlier version of this paper at a research seminar at IESEG School of Management. We appreciate the insightful comments from the participants, particularly Frank de Bakker.

## About the Authors

**Kaveh Mohajeri** is an associate professor in the Department of Innovation, Entrepreneurship, and Information Systems at IESEG School of Management, France. He received his BSc degree in industrial management and his MSc in IT management from the University of Tehran, Iran. He also has a PhD in information systems from Virginia Commonwealth University. A core part of his research agenda concerns advancing novel methodological thinking and research practices, focusing on the issues of research relevance and impact, deductive theorizing, and computationally intensive qualitative research. His work has been published in *MIS Quarterly*, *Journal of the Association for Information Systems*, and *Journal of Information Technology*, among other outlets.

**Amir Karami** is an Associate Professor of Business Analytics/Quantitative Methods in the Management, Information Systems, and Quantitative Methods department at Collat School of Business at the University of Alabama at Birmingham. His research interests include text mining, social media analytics, generative AI, health informatics, mis/disinformation, and computational social science. His work has been published in various journals, such as *International Journal of Information Management* and *Journal of Biomedical Informatics*.